\input epsf
\documentstyle[12pt,epsf]{article}
\newlength{\pubnumber} 

\catcode`\@=11
\@addtoreset{equation}{section}

\def\section{\@startsection{section}{1}{\z@}{3.5ex plus 1ex minus .2ex}
 {2.3ex plus .2ex}{\large\bf}}
\def\subsection{\@startsection{subsection}{2}{\z@}{2.3ex plus .2ex}
 {2.3ex plus .2ex}{\bf}}

\begin{document}

\begin{titlepage}
\samepage{
\setcounter{page}{1}
\rightline{UFIFT--HEP--96--16}
\rightline{\tt hep-ph/9607296}
\rightline{July 1996}
\vfill
\begin{center}
 {\Large \bf  A Low Energy Dynamical SUSY Breaking\\ Scenario\\
              Motivated from Superstring Derived Unification}
\vfill
\vfill
 {\large Alon E. Faraggi$^{1,2}$\footnote{
        E-mail address: faraggi@phys.ufl.edu}\\}
\vspace{.12in}
 {\it $^{1}$   Institute for Fundamental Theory, Department of Physics, \\
        University of Florida, Gainesville, FL 32611,
        USA\footnote{Permanent address.}\\}
\vspace{.075in}
 {\it  $^{2}$ CERN, Theory Division, \\
              1211 Geneva, Switzerland }
\end{center}
\vfill
\begin{abstract}
  {\rm
Recently there has been a resurgence of interest in gauge mediated
dynamical supersymmetry breaking scenarios. I investigate
how low energy dynamical SUSY breaking may arise from superstring
models. In a three generation string derived model I propose that
the unbroken hidden non--Abelian gauge group at the string scale
is $SU(3)_H$ with matter multiplets. Due to the small gauge content
of the hidden gauge group the supersymmetry breaking scale may be
consistent with the dynamical SUSY breaking scenarios.
The messenger states are obtained in the superstring model from sectors
which arise due to the ``Wilson--line'' breaking of the unifying
non--Abelian gauge symmetry. An important property of the
string motivated messenger states is the absence of superpotential terms
with the Standard Model states. The stringy symmetries therefore
forbid the flavor changing processes which may arise due to couplings
between the messenger sector states and the Standard Model states.
Motivated from the problem of string gauge
coupling unification I contemplate a scenario in which the messenger
sector consists solely of color triplets. This hypothesis predicts
a chargino mass below the $W$--boson mass.
Imposing the current limits
from the LEP1 and LEP1.5 experiments the lightest supersymmetric
particles predicted by this hypothesis are in the mass ranges
$M_{\chi^\pm}\approx55-65$ GeV,  $M_{\chi^0}\approx35-50$ GeV
and $M_{\tilde\nu}\approx45-60$ GeV which will be tested in the
forthcoming LEP2 experiments. }
\end{abstract}
\smallskip}
\end{titlepage}

\setcounter{footnote}{0}

\def\beq{\begin{equation}}
\def\eeq{\end{equation}}
\def\beqn{\begin{eqnarray}}
\def\eeqn{\end{eqnarray}}

\def\ie{{\it i.e.}}
\def\eg{{\it e.g.}}
\def\half{{\textstyle{1\over 2}}}
\def\third{{\textstyle {1\over3}}}
\def\quarter{{\textstyle {1\over4}}}
\def\m{{\tt -}}
\def\p{{\tt +}}

\def\slash#1{#1\hskip-6pt/\hskip6pt}
\def\slk{\slash{k}}
\def\GeV{\,{\rm GeV}}
\def\TeV{\,{\rm TeV}}
\def\y{\,{\rm y}}
\def\SM{Standard-Model }
\def\SUSY{supersymmetry }
\def\SSSM{supersymmetric standard model}
\def\vev#1{\left\langle #1\right\rangle}
\def\l{\langle}
\def\r{\rangle}

\def\Htw{{\tilde H}}
\def\chibar{{\overline{\chi}}}
\def\qbar{{\overline{q}}}
\def\ibar{{\overline{\imath}}}
\def\jbar{{\overline{\jmath}}}
\def\Hbar{{\overline{H}}}
\def\Qbar{{\overline{Q}}}
\def\abar{{\overline{a}}}
\def\alphabar{{\overline{\alpha}}}
\def\betabar{{\overline{\beta}}}
\def\tautwo{{ \tau_2 }}
\def\thetatwo{{ \vartheta_2 }}
\def\thetathree{{ \vartheta_3 }}
\def\thetafour{{ \vartheta_4 }}
\def\ttwo{{\vartheta_2}}
\def\tthree{{\vartheta_3}}
\def\tfour{{\vartheta_4}}
\def\ti{{\vartheta_i}}
\def\tj{{\vartheta_j}}
\def\tk{{\vartheta_k}}
\def\calF{{\cal F}}
\def\smallmatrix#1#2#3#4{{ {{#1}~{#2}\choose{#3}~{#4}} }}
\def\ab{{\alpha\beta}}
\def\Minv{{ (M^{-1}_\ab)_{ij} }}
\def\bone{{\bf 1}}
\def\ii{{(i)}}
\def\V{{\bf V}}
\def\b{{\bf b}}
\def\N{{\bf N}}
\def\t#1#2{{ \Theta\left\lbrack \matrix{ {#1}\cr {#2}\cr }\right\rbrack }}
\def\C#1#2{{ C\left\lbrack \matrix{ {#1}\cr {#2}\cr }\right\rbrack }}
\def\tp#1#2{{ \Theta'\left\lbrack \matrix{ {#1}\cr {#2}\cr }\right\rbrack }}
\def\tpp#1#2{{ \Theta''\left\lbrack \matrix{ {#1}\cr {#2}\cr }\right\rbrack }}
\def\l{\langle}
\def\r{\rangle}


\def\inbar{\,\vrule height1.5ex width.4pt depth0pt}

\def\IC{\relax\hbox{$\inbar\kern-.3em{\rm C}$}}
\def\IQ{\relax\hbox{$\inbar\kern-.3em{\rm Q}$}}
\def\IR{\relax{\rm I\kern-.18em R}}
 \font\cmss=cmss10 \font\cmsss=cmss10 at 7pt
\def\IZ{\relax\ifmmode\mathchoice
 {\hbox{\cmss Z\kern-.4em Z}}{\hbox{\cmss Z\kern-.4em Z}}
 {\lower.9pt\hbox{\cmsss Z\kern-.4em Z}}
 {\lower1.2pt\hbox{\cmsss Z\kern-.4em Z}}\else{\cmss Z\kern-.4em Z}\fi}

\def\AEF{A.E. Faraggi}
\def\NPB#1#2#3{{\it Nucl.\ Phys.}\/ {\bf B#1} (19#2) #3}
\def\PLB#1#2#3{{\it Phys.\ Lett.}\/ {\bf B#1} (19#2) #3}
\def\PRD#1#2#3{{\it Phys.\ Rev.}\/ {\bf D#1} (19#2) #3}
\def\PRL#1#2#3{{\it Phys.\ Rev.\ Lett.}\/ {\bf #1} (19#2) #3}
\def\PRT#1#2#3{{\it Phys.\ Rep.}\/ {\bf#1} (19#2) #3}
\def\MODA#1#2#3{{\it Mod.\ Phys.\ Lett.}\/ {\bf A#1} (19#2) #3}
\def\IJMP#1#2#3{{\it Int.\ J.\ Mod.\ Phys.}\/ {\bf A#1} (19#2) #3}
\def\nuvc#1#2#3{{\it Nuovo Cimento}\/ {\bf #1A} (#2) #3}
\def\etal{{\it et al\/}}

\hyphenation{su-per-sym-met-ric non-su-per-sym-met-ric}
\hyphenation{space-time-super-sym-met-ric}
\hyphenation{mod-u-lar mod-u-lar--in-var-i-ant}


\setcounter{footnote}{0}

Recently there has been renewed interest in the dynamical SUSY breaking
scenarios \cite{dsbearlypapers,dsb}.
In these scenarios supersymmetry breaking is generated dynamically
at a relatively low scale and is transmitted to the observable sector
by the gauge interactions of the Standard Model. An important property
of this type of gauge mediated supersymmetry breaking is the
natural suppression of flavor changing neutral currents. In these scenarios
the universality of the Standard Model gauge interactions results
in generation blind mass parameters for the supersymmetric scalar spectrum.
This attractive property of gauge mediated supersymmetry breaking
is an important advantage over some other possible scenarios.
A crucial assumption in this regard is the absence of interaction
terms between the messenger sector states and the Standard Model states.
In this paper I show that the absence of such interaction terms arises
naturally in string derived models.

In the dynamical gauge mediated SUSY breaking scenarios, supersymmetry
is broken nonperturbatively and the breaking is mediated to the observable
sector by a messenger sector. The messenger sector typically consists of
vector--like color triplets and electroweak doublets, beyond the spectrum
of the Minimal Supersymmetric Standard Model (MSSM).
The messenger sector states and their charges under the Standard Model
gauge group then determine the superparticle mass spectrum.

The gaugino masses are obtained from one--loop diagrams and are given by

\begin{equation}
M_i(\Lambda)={{\alpha_i(\Lambda)}\over {4\pi}}\Lambda
\label{gauginomasses}
\end{equation}
where $\Lambda$ is the SUSY breaking scale and $\alpha_i(\Lambda)$ are
the Standard Model coupling constants at the scale $\Lambda$.
The scalar masses arise from two--loop diagrams and are given by
\begin{equation}
m^2(\Lambda)=
2\Lambda^2\left\{C_3\left[{{\alpha_3(\Lambda)}\over{4\pi}}\right]^2
                +C_2\left[{{\alpha_2(\Lambda)}\over{4\pi}}\right]^2
                +{3\over5}\left({Y\over2}\right)^2
                             \left[{{\alpha_1(\Lambda)}\over{4\pi}}\right]^2
\right\}
\label{scalarmasses}
\end{equation}
where the weak hypercharge has the standard $SO(10)$ normalization
$U(1)_Y=3/5 U(1)_1$ and $C_3=4/3$ for color triplet scalars and zero
for sleptons and $C_2=3/4$ for electroweak doublets and zero for singlets.

In this paper I examine how a low--energy gauge--mediated dynamical SUSY
breaking scenario may arise from superstring derived models.
Traditionally it has been assumed that the supersymmetry breaking
in superstring models is generated dynamically by a hidden gauge group
with a large gauge content, typically $E_8$, $SO(10)$ or $SU(5)$
\cite{susyx}.
I propose that alternative scenarios exist in which the hidden
gauge group is broken at the Planck scale to a group with a small gauge
content, like $SU(3)$. The appearance of a ``small'' hidden gauge group
may result in the nonperturbative SUSY breaking dynamics at a hierarchically
low scale. This illustrates that the SUSY breaking dynamics may indeed be
generated at a relatively low scale in accordance with the gauge
mediated SUSY breaking scenarios. In this paper the SUSY breaking sector
will not be investigated in detail. Rather, following ref. \cite{FH},
some heuristic arguments are given which suggest that supersymmetry may
indeed be broken in the hidden sector.
Instead, the focus of this paper is on a predictive string--motivated
scenario with regard to the messenger sector.
In this scenario the messenger sector consists solely
of color triplets. There are no electroweak doublets in the messenger sector.
This scenario differs from previously studied dynamical SUSY breaking
scenarios in the context of unified SUSY models and results in specific
predictions with regard to the supersymmetric spectrum. Specifically,
in the simplest scenario
the lightest chargino is predicted to be below the $W$-boson mass, the
lightest neutralino is predicted to be of the order $35-50$ GeV and the
lightest scalar superpartner is the sneutrino with a mass of the order
$45-60$ GeV.

The messenger sector states are obtained in the superstring model
from sectors which arise due to the ``Wilson--line'' breaking
of the unifying non--Abelian gauge symmetry. A consequence of the
gauge symmetry breaking in superstring models by ``Wilson--line''
is the appearance of massless states which do not fit
into multiplets of the original unbroken gauge symmetry.
I refer to such states generically as exotic ``Wilsonian'' matter
states. This is an important property as it may result is
conserved quantum numbers which forbid the couplings of the
exotic ``Wilsonian'' matter states to the Standard Model states.
This is precisely what happens in the superstring derived model
under consideration. The Standard Model states are obtained
from three 16 multiplets of $SO(10)$ and the messenger sector
states are ``Wilsonian'' matter states. In this string derived
model it has been shown, to all orders of nonrenormalizable terms,
that there are no superpotential terms between the ``Wilsonian''
color triplets and the Standard Model states \cite{ps,ccf}.
This result illustrates that string models can indeed produce the
symmetries needed to prevent the interaction of the messenger sector
states with the Standard Model states. This demonstrates that, indeed,
string motivated gauge mediated SUSY breaking scenarios
can resolve the problematic supersymmetric contributions to
flavor changing neutral currents.

The motivation to consider a messenger sector with color triplets only
arises from the problem of string gauge coupling unification.
While in the Minimal Supersymmetric Standard Model (MSSM)
the gauge coupling are observed to intersect at a scale of the order
of $2\times 10^{16}$ GeV \cite{gcumssm},
string theory predicts that the unification scale
is of the order of $g_{\rm string}\times 5\times 10^{17}$ GeV \cite{scales},
with $g_{\rm string}\approx0.8$ at the unification scale.
Thus, approximately a factor of twenty separates the MSSM and
string unification scales.
It would seem that this discrepancy should have many possible resolutions,
keeping in mind that the gauge parameters are extrapolated over fifteen
orders of magnitude. Indeed, in string models there are many possible
sources that may affect the gauge coupling unification. For instance,
heavy string threshold corrections, light SUSY threshold corrections,
enhanced gauge group structure at an intermediate mass scale,
weak hypercharge normalization which differs from the standard GUT
normalization, intermediate matter thresholds and nonperturbative effects.
Surprisingly, however, the problem is not easily resolved. In ref. \cite{DF}
the string gauge coupling problem was analysed in the context of the
realistic free fermionic models. It was shown, in a wide range of
realistic free fermionic models, that heavy string threshold correction,
non--standard hypercharge normalizations, light SUSY thresholds or
intermediate gauge structure do not resolve the problem. Instead,
the problem may only be resolved due to the existence of additional
intermediate matter thresholds, beyond the MSSM \cite{gcu,gaillard,DF}.
This additional matter consists of color triplets and electroweak doublets,
in vector--like representations. Remarkably, some string models contain
in their massless spectrum the additional states, with the specific weak
hypercharge assignment needed to achieve string scale unification \cite{gcu}.
In ref. \cite{gcu,DF} it was shown that there exist many possible
scenarios for the mass scales of the additional color triplets and the
electroweak doublets. In general, to accommodate the string unification scale
with the experimental values for $\alpha_{\rm strong}(M_Z)$ and
$\sin^2\theta_W(M_Z)$ the additional vector--like color triplets
have to be much lighter than the additional electroweak doublets \cite{DF}.
Furthermore, the mass scale of the color triplets which is required
for them to play the role of the messenger sector in the dynamical SUSY
breaking scenarios, can also be compatible with the mass scale which is
needed to resolve the string gauge coupling unification problem.
It was recently also suggested that massive color triplets at the required
mass scale, $\Lambda\approx100$ TeV,
are also good dark matter candidates \cite{ccf}.
Thus, the same ``Wilsonian'' matter states that can provide the missing dark
matter, can also play the role of the messenger sector states in the
superstring motivated gauge mediated dynamical SUSY breaking scenario.
It is important to note that the superstring symmetries which
forbid the interaction of the ``Wilsonian'' matter states with the
Standard Model states, insure both their stability as well as the
absence of flavor mediating interactions from the messenger sector.

In this paper I propose in a specific superstring derived
standard--like model that a $SU(3)_H$ gauge group may be the
only non--Abelian part of the hidden gauge group which is left
unbroken at the string scale. The hidden $SU(3)_H$ gauge
group becomes strongly interacting at a relatively low scale.
Gaugino and matter condensation may then drive a non--vanishing F--terms
for one of the gauge singlet fields in the massless string spectrum.
This singlet couples to the messenger sector. In this paper motivated
from the problem of string--scale gauge coupling unification
I make the hypothesis that the messenger sector consists only
of color triplets while the electroweak doublets are much heavier.
The hypothesis that the
messenger sector consists solely of color triplets has crucial implications.
It predicts the existence of a chargino which is lighter than the
$W$--boson mass. Thus, this hypothesis will be confirmed or ruled out
in the forthcoming LEP2 experiments.

Examples of semi--realistic superstring models were constructed
in the orbifold and free fermionic
formulations \cite{ssm,rffm,eu,slm,gcu,custodial}.
The proposed scenario for superstring dynamical SUSY breaking is
illustrated in the superstring standard--like model of ref. \cite{gcu}.
The superstring derived standard--like models are constructed in the free
fermionic formulation \cite{FFF}. The realistic free fermionic models
\cite{rffm,eu,slm,gcu,custodial} are
defined in terms of a set of boundary condition basis vectors for all the
world--sheet fermions, and the one--loop GSO amplitudes. The physical spectrum
is obtained by applying the generalised GSO projections. The first five
basis vectors in the models that I consider
consist of the NAHE set, $\{{\bf 1},S,b_1,b_2,b_3\}$.
At the level of the NAHE set the observable gauge group is
$SO(10)\times SO(6)^3$ and the hidden gauge group is $E_8$. The
sectors $b_1$, $b_2$ and $b_3$ produce 48 generations in the 16
representation of $SO(10)$. Adding to the NAHE set three additional
boundary condition basis vectors, $\{\alpha,\beta,\gamma\}$
reduces the number of generations to three, one from each of the sectors
$b_1$, $b_2$ and $b_3$. At the same time the observable $SO(10)$ gauge group
is broken to one of its subgroups, the flavor $SO(6)^3$ symmetries are broken
to product of $U(1)$'s, and the hidden $E_8$ is broken to one of its subgroups.
It is important to note the correspondence between free fermionic models
and orbifold models. The free fermionic models correspond to $Z_2\times Z_2$
orbifold models with nontrivial background fields \cite{foc}.
The Neveu--Schwarz sector
corresponds the untwisted sector, and the sectors $b_1$, $b_2$ and $b_3$
correspond to the three twisted sectors of the $Z_2\times Z_2$ orbifold
models. The three sectors which break the $SO(10)$ symmetry correspond to
Wilson lines in the orbifold terminology.

In the superstring derived standard--like
models the Neveu--Schwarz sector gives rise to the generators of the
$SU(3)_C\times SU(2)_L\times U(1)_{B-L}\times U(1)_{T_{3_R}}\times U(1)^6$
gauge group in the observable sector. In the hidden sector the
Neveu--Schwarz sector produces the generators of the
$SU(5)\times SU(3)\times U(1)^2$ hidden gauge group.
The three sectors $b_1$, $b_2$ and $b_3$ produce three chiral 16
of $SO(10)$ decomposed under $SU(3)\times SU(2)\times U(1)^2$.
In addition to the spin one and two multiplets, the Neveu--Schwarz  (NS) sector
produces three pairs of electroweak doublets,
$\{h_1, h_2, h_3, {\bar h}_1, {\bar h}_2, {\bar h}_3\}$,
three pairs of $SO(10)$ singlets with $U(1)$ charges,
$\{\Phi_{12},\Phi_{23},\Phi_{13},{\bar\Phi}_{12},
{\bar\Phi}_{23}, {\bar\Phi}_{13}\}$,
and three singlets of the entire four dimensional gauge group,
$\{\xi_{1},\xi_2, \xi_3\}$.
The sector $b_1+b_2+\alpha+\beta$ produces one or two
additional Higgs pairs, $\{h_{45},{\bar h}_{45},h^\prime_{45},
{\bar h}^\prime_{45}\}$, and several $SO(10)$ singlet fields with horizontal
$U(1)$ charges,
$\{\Phi_{45},{\bar\Phi}_{45},\Phi^\prime_{45},{\bar\Phi}^\prime_{45},
\Phi_{1,2},{\bar\Phi}_{1,2}\}$.
The three sectors $b_j+2\gamma$ produce massless states
in the vector 16 representation of the $SO(16)$ subgroup of the hidden
$E_8$, decomposed under the final hidden gauge group,
$\{T_{1,2,3},{\bar T}_{1,2,3},V_{1,2,3},{\bar V}_{1,2,3}\}$.
The $T_i$ $({\bar T}_i)$ are 5 $({\bar 5})$ and the
$V_i$ $({\bar V}_i)$ are 3 $({\bar 3})$ of the hidden $SU(5)$ and $SU(3)$
gauge groups, respectively.
In addition, vectors that are combinations of the NAHE set basis vectors
and of the basis vectors $\{\alpha,\beta,\gamma\}$, produce additional
massless sectors which break the $SO(10)$ symmetry explicitly.
These are the sectors which arise due to the ``Wilson--line'' breaking
of the $SO(10)$ gauge symmetry.
In the model of ref. \cite{gcu} there are two pairs of additional
vector--like $SU(3)_C$ color triplets $\{D_1,D_2,{\bar D}_1,{\bar D}_2\}$
from the sectors $b_{1,2}+b_3+\beta\pm\gamma$.
These two color triplet pairs have the standard down--type weak hypercharge
assignment \cite{ccf}. However, they carry non--standard $SO(10)$ charges
under the $U(1)_{Z^\prime}$ which is embedded in $SO(10)$ and is orthogonal
to the weak hypercharge \cite{ccf}. In the model of ref. \cite{gcu},
because of the exotic charges of these color triplets under the
$U(1)_{Z^\prime}$ symmetry, there is a residual discrete symmetry
which forbids the couplings of these color triplets to the Standard Model
states \cite{ccf}. This is a crucial observation from which
follows that the interaction of these color triplets
with the Standard Model states is indeed only through the gauge interactions.
The stringy local discrete symmetries therefore forbid the couplings
which may result in flavor changing processes.
An additional pair of color triplets,
$\{D_3,{\bar D}_3\}$ with a non--standard weak hypercharge assignment
is obtained from the sector ${\bf1}+\alpha+2\gamma$. Three pairs of additional
electroweak doublets are obtained from the sectors
$\{1+b_i+\alpha+2\gamma\}$. The sectors $b_{1,2}+b_3+\beta\pm\gamma$
also produce two additional pairs of triplets of the hidden
$SU(3)_H$ gauge group. The total number of triplets of the hidden
$SU(3)_H$ gauge group in the model of ref. \cite{gcu} is ten.
Further details on the construction of the superstring
standard--like models and their spectrum are given in
ref. \cite{eu,slm,ccf}.

The cubic level and higher order nonrenormalizable terms in the
superpotential are obtained by calculating correlators between
vertex operators,
$A_N\sim\langle V_1^fV_2^fV_3^b\cdot\cdot\cdot V_N^b\rangle,$
where $V_i^f$ $(V_i^b)$ are the fermionic (scalar)
components of the vertex operators.
The non-vanishing terms must be invariant under all the symmetries of the
string models and must satisfy all the string selection rules \cite{kln}.
A detailed analysis of texture of fermion mass matrices was done in
refs. \cite{nrt,ckm} for the model of ref. \cite{eu}. The analysis
was done up to nonrenormalizable terms of order $N=8$.
{}From this analysis
it was found that the two sectors $b_1$ and $b_2$ produce the two heavy
generations while the sector $b_3$ produces the lightest generation.
The mixing terms between the generations are obtained by exchanging
states which transform under the 16 vector representation of the
hidden $SO(16)$ gauge group. For example,
the lowest order mixing terms in the model of ref. \cite{eu} arise
at order $N=6$,
\begin{eqnarray}
&&d_3Q_2h_{45}\Phi_{45}V_3{\bar V_2}~~~~~~~~~~~
  d_2Q_3h_{45}\Phi_{45}V_2{\bar V_3},\nonumber\\
&&d_3Q_1h_{45}\Phi_{45}{\bar V}_1V_3~~~~~~~~~~~
  d_1Q_3h_{45}\Phi_{45}{V}_1{\bar V}_3.
\label{mixingtermsin278}
\end{eqnarray}
where $V_i$, ${\bar V}_i$ transform as $3$ and ${\bar 3}$ of the
hidden $SU(3)_H$ gauge group. Thus, in this model in order to obtain
a phenomenologically acceptable Cabbibo angle the hidden $SU(3)_H$
gauge group has to be broken \cite{ckm}. In that case the nonperturbative
dynamics in the hidden $SU(5)_H$ gauge group generate gaugino
and matter condensation that may break supersymmetry with
$m_{3/2}\sim1$ TeV \cite{FH}. Thus, the supersymmetry breaking
scenario in this model is similar to supersymmetry breaking
in the traditional supergravity models.

The structure of the model of ref. \cite{gcu} is similar to the structure
of the model of ref. \cite{eu}. In particular, the observable spectrum
from the NS sector and the sectors $b_1$, $b_2$ and $b_3$ is
identical in the two models with some differences in the
charges under the horizontal $U(1)$ symmetries, $U(1)_{{L,R}_{4,5,6}}$.
The model of ref. \cite{gcu} produces similarly potential mass terms
for the two heavy generations from quartic and quintic order terms.
A detailed analysis of the texture of fermion mass matrices in not
the purpose of the present paper. For our purposes here, it is sufficient to
note that in the model or ref. \cite{gcu} the generation
mixing terms are obtained at
order $N=6$
\begin{eqnarray}
&&d_3Q_2h_{45}^\prime\Phi_{45}T_2{\bar T}_3~~~~~~~~~~~
  d_3Q_1h_{45}^\prime\Phi_{45}T_1{\bar T}_3\nonumber\\
&&d_2Q_3h_{45}^\prime\Phi_{45}{\bar T}_2T_3~~~~~~~~~~~
  d_1Q_3h_{45}^\prime\Phi_{45}{\bar T}_1T_3\label{mixingtermsin302}
\end{eqnarray}
where $T_i$ and ${\bar T}_i$ transform as $5$ and ${\bar 5}$
of the hidden $SU(5)_H$ gauge group. Thus, in this model in order
to generate a Cabbibo angle of a phenomenologically acceptable order
of magnitude, the hidden $SU(5)$ gauge group has to be broken while
the hidden $SU(3)_H$ gauge group is left unbroken. The hidden $SU(5)_H$
gauge group can of course be broken in several possible patterns. I will
assume that there exist a solution in which it is completely broken.
In this case the SUSY breaking dynamics will be driven by the hidden
$SU(3)_H$ gauge group.

Thus, in the model of ref. \cite{gcu}, to obtain a sizable generation mixing
the hidden gauge group needs to be broken, and only $SU(3)_H$ remains
unbroken. In this case because of the small gauge content of the
$SU(3)$ gauge group, the scale where the hidden $SU(3)_H$ gauge
group becomes strongly interacting can be much smaller, in accordance
with the gauge mediated dynamical SUSY breaking scenarios.
The scale at which the hidden $SU(3)_H$ gauge group becomes
strongly interacting is given by
\begin{equation}
\Lambda_3=M_{\rm string}~
{\rm exp}{({{2\pi}\over{b}}{{(1-\alpha_0)}\over\alpha_0})},
\label{sisu3}
\end{equation}
where $M_{\rm string}$ is the string unification scale,
$b=1/2~n_3-9$ is the $\beta$--function coefficient
of the hidden $SU(3)_H$ gauge group and $\alpha_0$ is the
gauge coupling constant at the string unification scale.
The scale at which the hidden $SU(3)$ gauge group becomes
strongly interacting depends on the $M_{\rm string}$, $\alpha_0$
and on the number of hidden $SU(3)_H$ triplets which are massless
at the string scale. For example, with $M_{\rm string}=4\times10^{17}$ GeV,
$\alpha_0=1/24$ and $n_3=8$, gives $\Lambda_3\approx100$ TeV.
This is roughly the scale required in the dynamical SUSY
breaking scenarios to obtain phenomenologically viable gaugino masses.
This illustrates that dynamical low energy SUSY breaking may indeed
be generated from superstring derived models. Detailed scenarios for the
mass scales of the hidden $SU(3)_H$ matter states can be studied from
an analysis of nonrenormalizable terms in the superpotential.

The superstring model under consideration contains in its massless spectrum
two pairs of color triplets $\{D_1,D_2,{\bar D}_1,
{\bar D}_2\}$ from the sectors $b_{1,2}+b_3+\beta\pm\gamma$.
with the charge assignment $({\bar 3},1)_{1/3}$, $(3,1)_{-1/3}$,
and one pair, $\{D_3,{\bar D}_3\}$ from the
sector $1+\alpha+2\gamma$ with charges $({\bar 3},1)_{1/6}$,
$(3,1)_{-1/6}$, under $SU(3)_C\times SU(2)_L\times U(1)_Y$.
The first two pairs transform as regular down--type quarks under the
Standard Model gauge group. The cubic level superpotential in this model
is given by,
\begin{eqnarray}
W_3&&=\{(
{u_{L_1}^c}Q_1{\bar h}_1+{N_{L_1}^c}L_1{\bar h}_1+
{u_{L_2}^c}Q_2{\bar h}_2+{N_{L_2}^c}L_2{\bar h}_2+
{u_{L_3}^c}Q_3{\bar h}_3+{N_{L_3}^c}L_3{\bar h}_3)\nonumber\\
&&
+{{h_1}{\bar h}_2{\bar\Phi}_{12}}
+{h_1}{\bar h}_3{\bar\Phi}_{13}
+{h_2}{\bar h}_3{\bar\Phi}_{23}
+{\bar h}_1{h_2}{\Phi_{12}}
+{\bar h}_1{h_3}{\Phi_{13}}
+{\bar h}_2{h_3}{\Phi_{23}}\nonumber\\
&&
+\Phi_{23}{\bar\Phi}_{13}{\Phi}_{12}
+{\bar\Phi}_{23}{\Phi}_{13}{\bar\Phi}_{12}
+{\bar\Phi}_{12}({\bar\Phi}_1{\bar\Phi}_1
+{\bar\Phi}_2{\bar\Phi}_2)
+{{\Phi}_{12}}(\Phi_1\Phi_1
+\Phi_2\Phi_2)\nonumber\\
&&
+{1\over2}\xi_3(\Phi_{45}{\bar\Phi}_{45}
+h_{45}{\bar h}_{45}+{\Phi_{45}^\prime}{{\bar\Phi}_{45}^\prime}
+h_{45}^\prime{\bar h}_{45}^\prime
+\Phi_1{\bar\Phi}_1+\Phi_2{\bar\Phi}_2)\nonumber\\
&&
+h_3{\bar h}_{45}{\bar\Phi}_{45}^\prime
+{\bar h}_3h_{45}{\Phi}_{45}^\prime
+h_3{\bar h}_{45}^\prime\Phi_{45}
+{\bar h}_3h_{45}^\prime{\bar\Phi}_{45}\nonumber\\
&&
+{1\over2}(\xi_1D_1{\bar D}_1+\xi_2D_2{\bar D}_2)
+{1\over\sqrt{2}}(D_1{\bar D}_2\phi_2+{\bar D}_1D_2{\bar\phi}_1)\}
\label{cubiclevel}
\end{eqnarray}
The color triplet pairs $\{D_1,D_2,{\bar D}_1,
{\bar D}_2\}$ can serve as the messenger sector states in the
superstring model and
the terms in the cubic level superpotential
${1\over2}(\xi_1D_1{\bar D}_1+\xi_2D_2{\bar D}_2)
+{1\over\sqrt{2}}(D_1{\bar D}_2\phi_2+{\bar D}_1D_2{\bar\phi}_1)$
can serve as the coupling between the messenger sector and the SUSY breaking
sector, provided that a non--vanishing $F$--term is generated in the
$\xi_{1,2}$ or $\phi_{1,2}$ directions. A detailed analysis of the
matter and gaugino condensates was performed in ref. \cite{FH}
for the model of ref. \cite{eu}. There indeed it was argued that
a non--vanishing $F$--term  in the $\xi_i$ direction is generated
due to the hidden matter condensates.
There, \cite{FH}, it was argued that the flat $F$ directions of the
cubic level superpotential are lifted once the nonrenormalizable terms
which include the hidden sector matter condensates are included in the
analysis. Again it should be emphasised
that the analysis was carried out in detail for the model of ref. \cite{eu}
where the hidden $SU(5)_H$ is left unbroken at the string scale.
However, due to the similar structure of the spectrum of the two models
and in particular the almost identical spectrum from the NS sector
and the sectors $b_j+2\gamma$, similar results are also expected to
hold in the case of the model of ref. \cite{gcu}.
Here I am assuming that indeed such a non--vanishing $F$--term
is generated by nonperturbative effects
in the $\xi$ direction and a more detailed analysis is left for
future work. With this assumption the color triplet pairs
$\{D_1, D_2,{\bar D}_1, {\bar D}_2\}$ serve as the messenger sector.
The messenger sector therefore consists solely of color triplets.
This implies specific predictions for the supersymmetric mass spectrum
which are studied below.

The gaugino and Higgsino mass spectrum is obtained by diagonalizing the
chargino and neutralino mass matrices.
The chargino mass matrix is given by
\begin{equation}
M_{\tilde C}=\left(\matrix{{\tilde
M}_2&M_W\sqrt{2}\sin\beta\cr
                             M_W\sqrt{2}\cos\beta&\mu\cr}\right)~,
\label{chargino}
\end{equation}
and the neutralino mass matrix is given by
\begin{equation}
M_{\tilde N}=\left(\matrix{
{\tilde M}_1&0&-M_Z{\sin\theta_W}{\cos\beta}&M_Z{\sin\theta_W}{\sin\beta} \cr
0&{\tilde M}_2&M_Z{\cos\theta_W}{\sin\beta}&-M_Z{\cos\theta_W}{\cos\beta} \cr
-M_Z{\sin\theta_W}{\cos\beta}&M_Z{\cos\theta_W}{\sin\beta}&0&\mu \cr
M_Z{\sin\theta_W}{\sin\beta}&-M_Z{\cos\theta_W}{\cos\beta}&\mu&0 \cr}\right),
\label{neutralino}
\end{equation}
With the assumption that the messenger sector consists only of
color triplets, it follows from Eq. \ref{gauginomasses} that
\begin{equation}
{\tilde M}_2=0,
\label{m2}
\end{equation}
in the chargino and neutralino mass matrices, Eqs. \ref{chargino} and
\ref{neutralino}, respectively. It follows from this hypothesis that the
lightest chargino is lighter than the $W$--boson mass. This is therefore
a precise prediction of the hypothesis that the messenger sector consists
solely of color triplets, which is motivated from the string gauge coupling
unification. This should be compared with the gauge mediated SUSY breaking
scenarios in the context of the MSSM. There one introduces a messenger
sector which consists of color triplets and electroweak doublets in
order not to spoil the intersection of the gauge couplings at the MSSM
unification scale.

To study the possible spectrum of the lightest
superparticles the two parameters in Eqs. \ref{chargino} and \ref{neutralino},
$\tan\beta$ and $\mu$ are varied in the ranges $1\le\tan\beta\le3$ and
$-50~{\rm GeV}\le\mu\le50~{\rm GeV}$ and with $\Lambda=100$ TeV. The current
limits on the chargino and neutralino masses from the LEP1 and LEP1.5
experiments are imposed in the analysis.

\begin{figure}[htb]
\centerline{\epsfxsize 3.0 truein \epsfbox {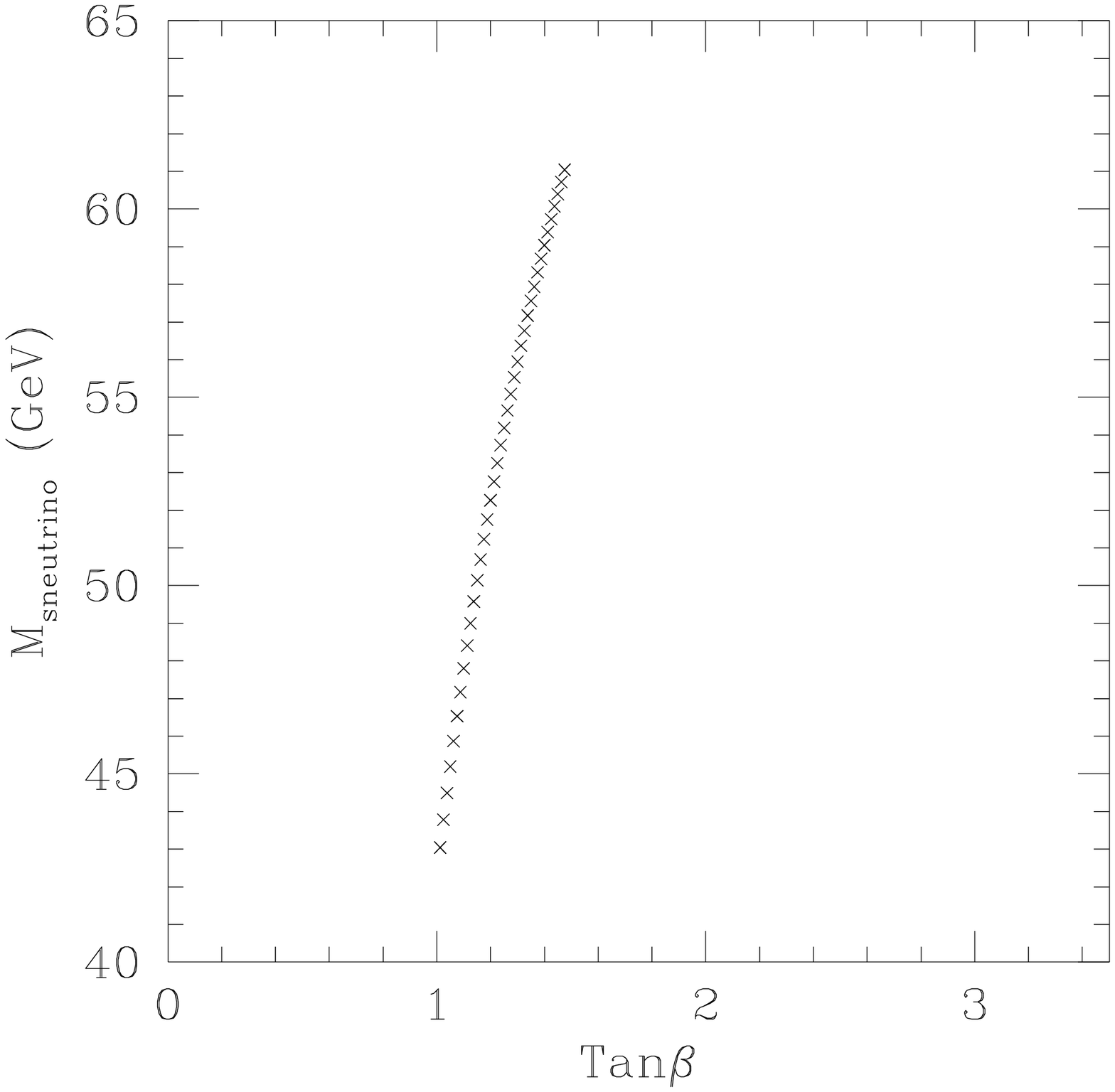}}
\caption{The predicted value of the lightest scalar superparticle versus
$\tan\beta$, with $1\le\tan\beta\le3$, $-50~{\rm GeV}\le\mu\le50~{\rm GeV}$ and
$\Lambda_3=100~{\rm TeV}$.}
\label{tbsn}
\end{figure}

Using the LEP1 and LEP1.5 data the four LEP experiments have imposed strong
constraints on the existence of charginos and neutralinos below the
$W$--boson mass \cite{lepexp,delphi}. These limits are often obtained
by making various assumptions on GUT boundary conditions for the gaugino
masses and on the scalar mass spectrum. In particular if the sneutrino mass
is assumed to be larger than 200 GeV then the chargino mass is
constrained to be larger than $\approx65$ GeV. However,
if the assumption on the sneutrino mass is relaxed then destructive
interference between the $s$--channel exchange diagram of the $Z$ and $\gamma$
gauge bosons and the $t$--channel exchange diagram of the sneutrino allows
smaller chargino masses. A recent analysis was done by the DELPHI
collaboration \cite{delphi} from which one can infer the conservative
limits of $m_{\chi^\pm}\ge56$ GeV and $m_{\chi^0}\ge35$ GeV.
The light sneutrino region is precisely the scenario predicted by the
superstring motivated dynamical SUSY breaking scenario with a messenger
sector which consists only of color triplets. As can be seen from Eq.
\ref{scalarmasses} in this case the second term in Eq. \ref{scalarmasses}
is equal to zero and the only contribution to the sneutrino mass
arises from the term due to the weak hypercharge. In the
analysis of the sparticle masses I have included the $D$--term contribution
given by
\begin{equation}
d_{\tilde p}=2\left(T_{3_L}^{\tilde p}-{3\over5}Y^{\tilde p}
{\tan^2\theta_W}\right)\cos2\beta M_W^2,
\label{dterm}
\end{equation}
where $T_{3_L}^{\tilde p}$ and $Y^{\tilde p}$ are the $SU(2)_L$ and
$U(1)_Y$ charges of a sparticle. In fig. \ref{tbsn} the predicted
sneutrino mass is plotted versus the electroweak VEVs ratio $\tan\beta$
and the scale $\Lambda$ is taken at 100 TeV.

\begin{figure}[htb]
\centerline{\epsfxsize 3.0 truein \epsfbox {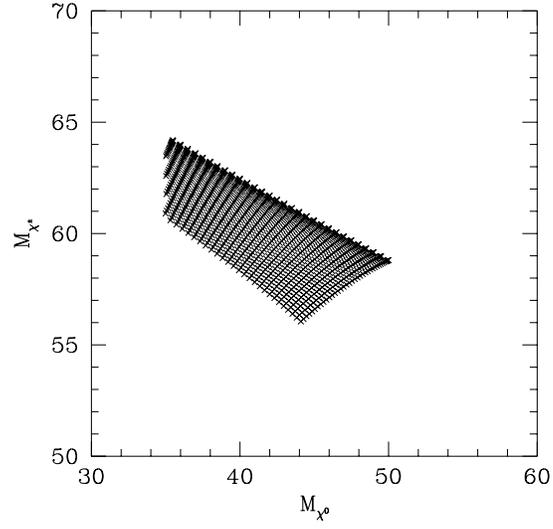}}
\caption{The predicted values of the lightest neutralino and the lightest
chargino for the same parameters values as in figure 1.}
\label{nech}
\end{figure}

In fig. \ref{nech}, the lightest chargino mass is plotted versus the
lightest neutralino mass, where the constraints $M_{\chi^0}\ge35$ GeV
and $M_{\chi^\pm}\ge56$ GeV have been
imposed. The predicted lightest neutralino mass is approximately
in the range $35-50~{\rm GeV}$ and the lightest chargino mass is
in the approximate range $56-65~{\rm GeV}$.
The next lightest neutralino, $\chi^1$, is found to be heavier than
53 GeV with $M_{\chi^0}+M_{\chi^1}>95$ GeV over the entire parameter
space. The next lightest and heaviest neutralinos, $\chi^2$ and $\chi^3$
are found with $M_{\chi^2}>84$ GeV and $M_{\chi^3}>165$ GeV and the
heavy chargino is found with $M^h_{\chi^\pm}>100$ GeV over the
parameter space.

\begin{figure}[htb]
\centerline{\epsfxsize 3.0 truein \epsfbox {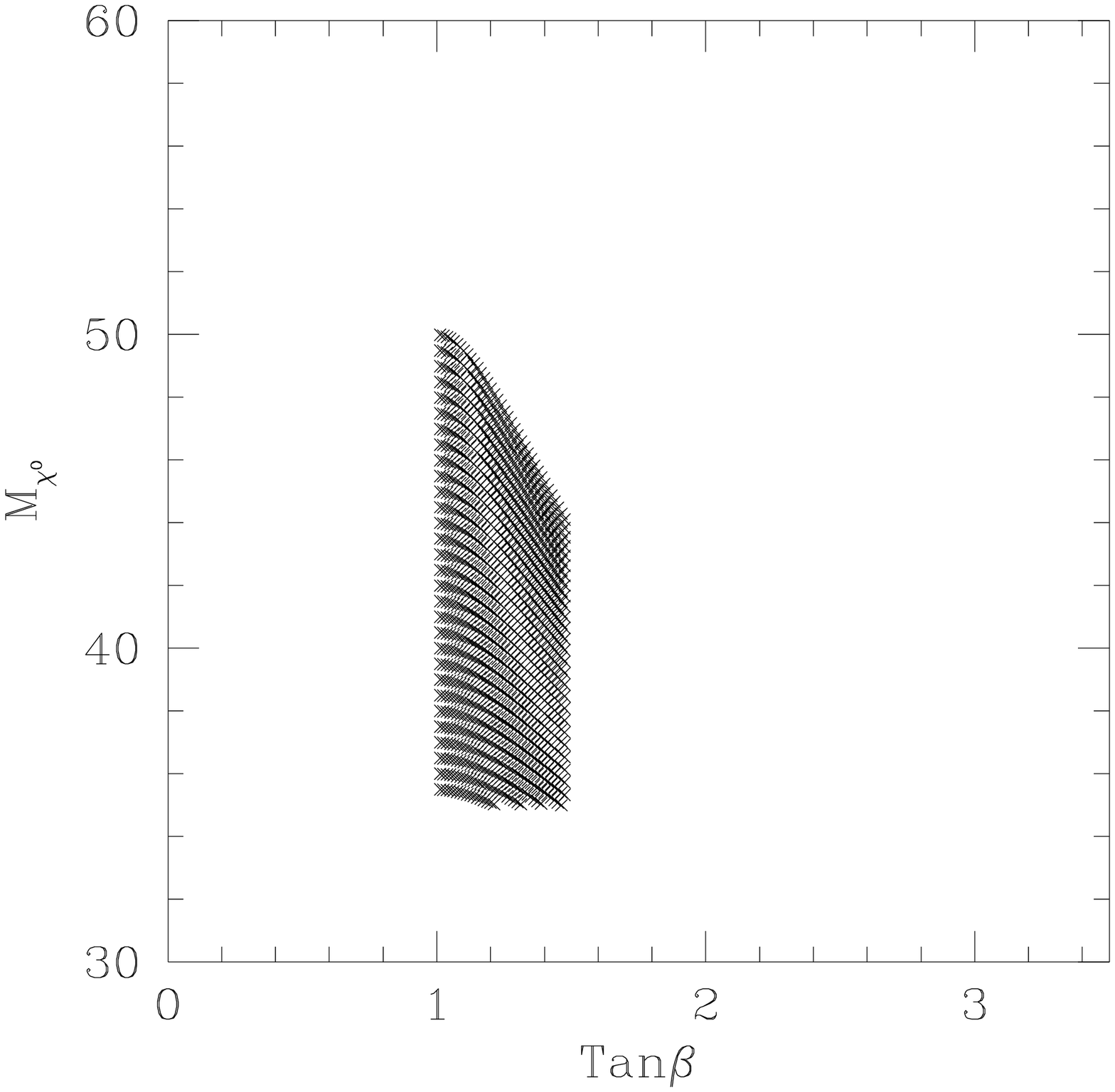}}
\caption{ The predicted value of the lightest neutralino mass versus
$\tan\beta$ the electroweak VEVs ratio,
for the same parameters values as in figure 1.}
\label{tbne}
\end{figure}

\begin{figure}[htb]
\centerline{\epsfxsize 3.0 truein \epsfbox {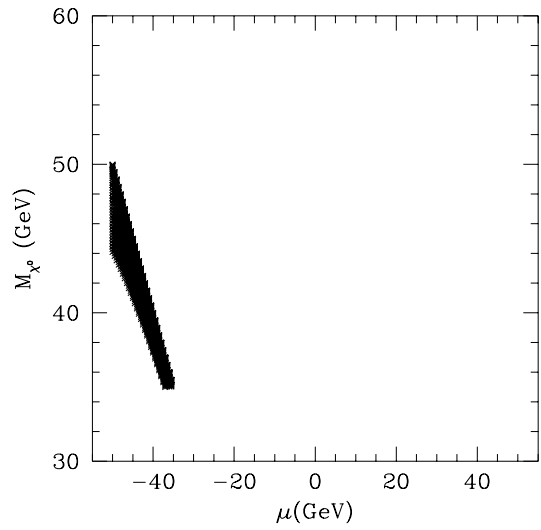}}
\caption{The predicted value of the lightest neutralino mass
versus $\mu$ the Higgs mixing parameter,
for the same parameters values as in figure 1.}
\label{mune}
\end{figure}

\begin{figure}[htb]
\centerline{\epsfxsize 3.0 truein \epsfbox {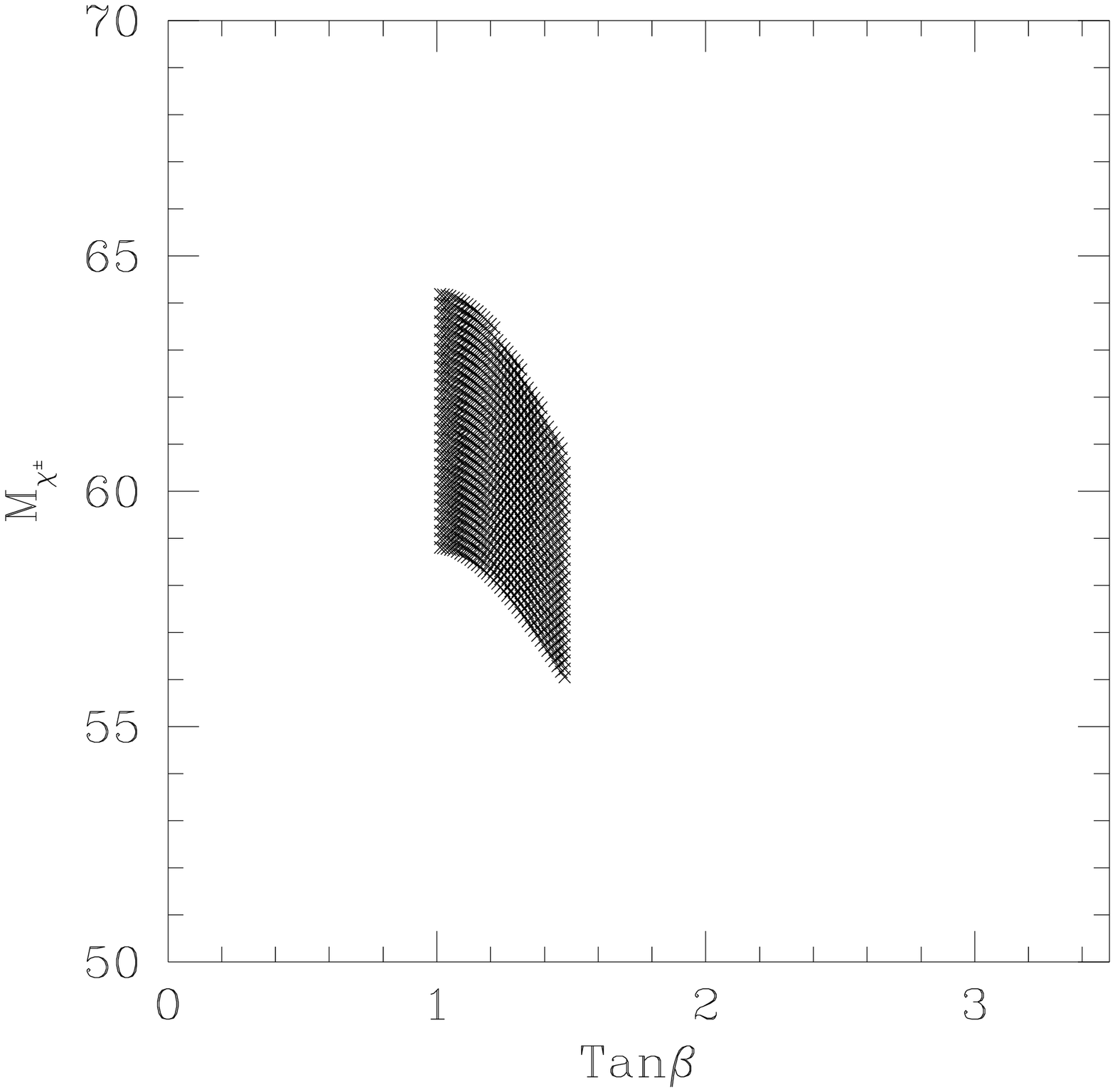}}
\caption{The predicted value of the lightest chargino mass versus
$\tan\beta$ the electroweak VEVs ratio,
for the same parameters values as in figure 1. }
\label{tbch}
\end{figure}

\begin{figure}[htb]
\centerline{\epsfxsize 3.0 truein \epsfbox {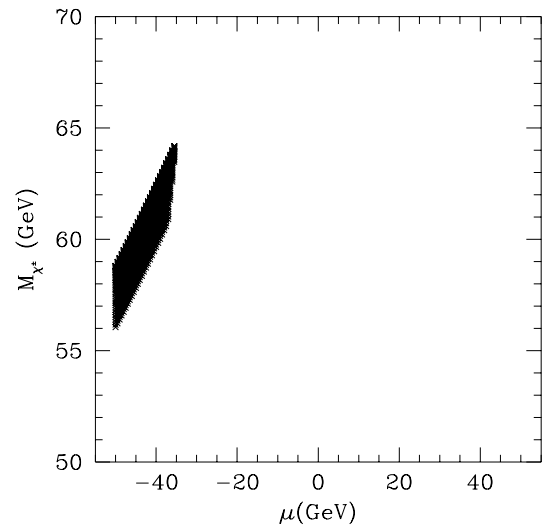}}
\caption{The predicted value of the lightest chargino mass versus
versus $\mu$ the Higgs mixing parameter,
for the same parameters values as in figure 1.}
\label{much}
\end{figure}

In figs. \ref{tbne} and \ref{mune} the predicted lightest neutralino mass is
shown versus the electroweak VEV ratio $\tan\beta$ and the Higgs mixing
parameter $\mu$, and in figs. \ref{tbch} and \ref{much} the predicted lightest
chargino mass is plotted versus the same parameters.
It is important to note that the hypothesis made in Eq. \ref{m2}
severely restricts the allowed parameter space for $\tan\beta$ and
$\mu$. It is seen from Eq. \ref{chargino}, with ${\tilde M}_2=0$,
that the absolute value of the $\mu$ parameter is constrained. With
${\tilde M}_2=0$ in Eq. \ref{chargino} $\mu$ acts as a seesaw scale and
as $\mu$ increases, the lightest chargino mass is pushed down. Thus, in this
scenario there is an upper limit on the allowed $\mu$ value. Similarly,
as can be seen from figs. \ref{tbne} and \ref{tbch} the small values for
$\tan\beta$, with $\tan\beta\approx1.1-1.4$ are preferred.

In this paper I examined how low energy dynamical SUSY breaking may arise
from superstring derived models. The gauge mediated SUSY breaking scenarios
have the important property of naturally suppressing supersymmetric
contributions to neutral flavor changing transitions.
In a specific superstring derived
standard--like model I have shown that requiring potentially realistic
fermion mass spectrum may result in a small hidden gauge group, like $SU(3)$.
In the case that the hidden gauge group is broken entirely to $SU(3)_H$
the nonperturbative hidden gauge dynamics may indeed result in
supersymmetry breaking at a low scale, in accordance with the gauge
mediated dynamical SUSY breaking scenarios. The messenger sector states
are obtained in the superstring model from sectors which arise due to the
``Wilson--line'' breaking of the $SO(10)$ gauge symmetry. As a result
the interaction terms with the Standard Model states are suppressed.
This illustrates that superstring models can provide the symmetries which
are needed to suppress the supersymmetric contributions to flavor changing
neutral currents. In this case the SUSY breaking is indeed transmitted
to the observable sector only by the Standard Model gauge interaction
which are generation blind. Thus, in the string inspired gauge mediated
dynamical SUSY breaking scenario the suppression of flavor changing neutral
currents arises naturally. It is also noted that the same
color triplet fields arise as stable states in the superstring models
and can be good dark matter candidate. Interestingly the mass scale
needed for this type of color triplets to be good dark matter candidates
is roughly the same as the mass scale for these states to serve as the
messenger sector in the dynamical SUSY breaking scenarios.
Motivated from the
problem of string gauge coupling unification I have made the hypothesis
that the messenger sector in the string motivated dynamical supersymmetry
breaking scenario consists solely of color triplets. This hypothesis
results in specific predictions for the superparticle spectrum
which will be tested in the forthcoming LEP2 experiments. In particular,
the lightest chargino mass is predicted to be below the $W$--boson mass.
More detailed analysis of the composition of the lightest chargino and
neutralino, which results from this hypothesis, and their experimental
signature is of further interest.

\bigskip
I would like to thank I. Antoniadis, S. Dimopoulos, G. Farrar
S. Thomas, F. Zwirner and especially J. Wells for important discussions,
and the CERN theory division for its hospitality while this work was conducted.
This work is supported in part by DOE Grant No.\ DE-FG-0586ER40272.

\vfill\eject

\bigskip
\medskip

\bibliographystyle{unsrt}

\vfill\eject

\end{document}